# Efficient near-infrared organic light-emitting diodes with emission from spin doublet excitons

Hwan-Hee Cho[1], Sebastian Gorgon[1], Giacomo Londi[2], Samuele Giannini[3], Changsoon Cho[1,4], Pratyush Ghosh[1], Claire Tonnelé[5], David Casanova[5,6], Yoann Olivier[2], Feng Li[7], David Beljonne[3], Neil C. Greenham[1,*], Richard H. Friend[1,*], Emrys W. Evans[8,*]

[1]Cavendish Laboratory, University of Cambridge, JJ Thomson Avenue, Cambridge, CB3 0HE, United Kingdom.

[2]Laboratory for Computational Modelling of Functional Materials, Namur Institute of Structured Matter, University of Namur, Rue de Bruxelles 61, 5000 Namur, Belgium.

[3]Laboratory for Chemistry of Novel Materials, University of Mons, 7000 Mons, Belgium.

[4]Department of Material Science and Engineering, Pohang University of Science and Technology (POSTECH), Pohang 37673, Republic of Korea

[5]Donostia International Physics Centre, Donostia, Euskadi, Spain.

[6]Ikerbasque Foundation for Science, 48009 Bilbao, Euskadi, Spain.

[7]State Key Laboratory of Supramolecular Structure and Materials, College of Chemistry, Jilin University, Changchun, China.

[8]Department of Chemistry, Swansea University, Singleton Park, Swansea, SA2 8PP, United Kingdom.

*Corresponding author.

Neil C. Greenham: Email: ncg11@cam.ac.uk

Richard H. Friend: E-mail: rhf10@cam.ac.uk

Emrys W. Evans: E-mail: emrys.evans@swansea.ac.uk

# Abstract

The development of luminescent organic radicals has resulted in materials with excellent optical properties for near-infrared (NIR) emission. Applications of light generation in this range span from bioimaging to surveillance. Whilst the unpaired electron arrangements of radicals enable efficient radiative transitions within the doublet-spin manifold in organic light-emitting diodes (OLEDs), their performance is limited by non-radiative pathways introduced in electroluminescence. Here, we present a host:guest design for OLEDs that exploits energy transfer with demonstration of up to 9.6% external quantum efficiency (EQE) for 800 nm emission. The tris(2,4,6-trichlorophenyl)methyl-triphenylamine (TTM-TPA) radical guest is energy-matched to the triplet state in a charge-transporting anthracene-derivative host. We show from optical spectroscopy and quantum-chemical modelling that reversible host-guest triplet-doublet energy transfer allows efficient harvesting of host triplet excitons.

# Main text

# Introduction

Advances in efficient near-infrared (NIR) organic light-emitting diodes (OLEDs) can enable light generation in the biological window for healthcare diagnosis and treatment. The requirement for long-wavelength light generation beyond the visible range is also motivated by communications and security applications. Whilst > 20% external quantum efficiency (EQE) in electroluminescence (EL) has been demonstrated for visible-light OLEDs, and commercial displays are commonplace, the performance of NIR OLEDs is generally limited to 5% EQE using fully-organic emitters with emission peak wavelengths at 800 nm and longer.[1] The materials approach and mechanisms for efficient visible-light OLEDs based on maximising luminescence from singlet and triplet excitons have not translated to efficient NIR OLEDs.

Doublet fluorescence from organic radicals is an emerging basis for highly efficient NIR light-emitting devices that exploit favourable optical, electronic and spin properties for optoelectronics.[2–17] Luminescent organic radicals enable high photoluminescence quantum yield (PLQY) in the NIR range, where immunity from normal 'energy gap law' considerations is linked to their unique electronic structure.[18] Almost 100% internal quantum efficiency (IQE) for EL was demonstrated in radical OLEDs exploiting tris(2,4,6-trichlorophenyl)methyl (TTM)-based radicals.[4] This performance shows that using the doublet-spin manifold in radicals for luminescence can circumvent typical efficiency limits (25% IQE) arising from the formation of singlet and triplet excitons in standard closed-shell molecule-based devices.[16,17] We recently reported efficient NIR OLEDs from triphenyl amine-substituted (2-chloro-3-pyridyl)bis(2,4,6-trichlorophenyl)methyl (TPA-PyBTM') with a maximum EQE of 6.4% for 800 nm peak emission.[19] However, these devices showed efficiency roll-off at high current densities and were limited by unbalanced electron and hole currents. Energy transfer mechanisms using thermally-activated delayed fldrvuorescence (TADF) materials for charge recombination and sensitization of radical emitters for EL showed promise for moving the exciton generation event away from radicals to combat the performance shortfall.[20]

Here, we use an anthracene derivative, 2-methyl-9,10-bis(naphthalene-2-yl)anthracene (MADN), as a host component that enables efficient charge transport to generate excitons that transfer to a triphenyl amine-substituted tris(2,4,6-trichlorophenyl)methyl (TTM-TPA) NIR radical emitter (see ref.[18] for synthetic details). The high-energy singlet state (near 3 eV) mitigates losses from the 'energy gap law' in MADN, whilst spin-allowed transfer between long-lived, low-energy host triplet and emissive radical doublet exciton states leads to efficient delayed emission. A maximum EQE for OLEDs of 9.6% is obtained at ~800 nm with reduced efficiency roll-off, enhanced radiance and device stability.

## Results and Discussion

**Near-infrared radical design of intersystem energy transfer**

Figure 1a shows the available energy transfer pathways between singlet ($S_1$) and triplet ($T_1$) excitons of MADN and doublet ($D_1$) excitons of TTM-TPA in the MADN:TTM-TPA system, and their chemical structures are depicted in Fig. 1b. The scheme shows the opportunity for energy harvesting of both singlet and triplet excitations in the non-radical host to form radical dopant states. This strategy exploits efficient spin-conserving transfer processes in the two pathways: singlet-doublet Förster resonance energy transfer (FRET) ($S_1 + D_0 \rightarrow S_0 + D_1$) and triplet-doublet Dexter energy transfer (DET) ($T_1 + D_0 \rightarrow S_0 + D_1$). MADN triplet emission extends between 700 nm and 900 nm,[21] which is energy-resonant with TTM-TPA doublet emission (Fig. 1c). Accordingly, the MADN:TTM-TPA system enables the study of exciton harvesting in the limit of small energy difference ($|\Delta E_{TD}| < 0.1$ eV) between MADN $T_1$ and TTM-TPA $D_1$, and where substantial host non-radiative losses due to the 'energy gap law' are minimised.

The steady-state photophysical properties of TTM-TPA, MADN, and 4,4'-bis(carbazol-9-yl)biphenyl (CBP) are depicted in Fig. 1c. Steady-state photophysical characteristics of TTM-TPA in different solutions are shown in Supplementary Fig. 1. The variation of photoluminescence (PL) peak wavelength between 730 nm and 890 nm with an increase in solvent polarity suggests the formation of intramolecular charge-transfer (CT) excited states, in line with our previous reports.[4,19] The CBP:TTM-TPA system is used as a reference for studies of energy transfer mechanisms in the MADN:TTM-TPA system. The PL spectra of MADN and CBP neat films overlap with the absorption of TTM-TPA so that photoexcited singlet excitons generated in the host non-radical components in MADN:TTM-TPA and CBP:TTM-TPA systems (at 330 nm for CBP and 370 nm for MADN: see Supplementary Fig. 2 for absorption spectra for CBP and MADN neat films) undergo efficient singlet-doublet transfer to the radical guest, with NIR fluorescence observed at ~800 nm (Fig. 1c). Small contributions of host emission to the total PL are observed and provide characteristic signatures for the singlet-doublet energy transfer channels in these systems. The PLQY of TTM-TPA in toluene is 24% (excited at 370 nm), while CBP:TTM-TPA 3% and MADN:TTM-TPA 3% films have 19% and 27% PLQY, respectively, at the same host excitation wavelength. This shows that high-energy singlet materials with efficient singlet-doublet transfer can be used to host NIR radical emitters.

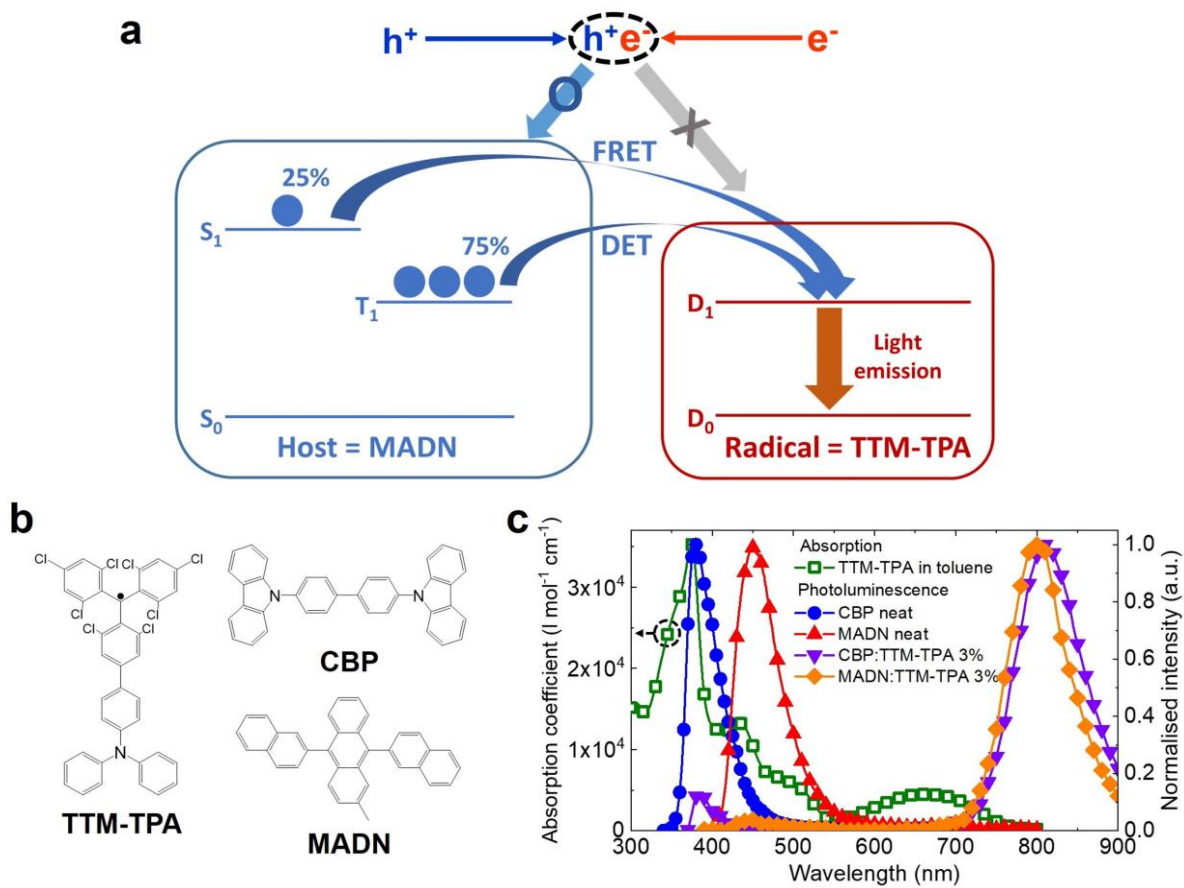

**Fig. 1 | Radical energy harvesting design for high efficiency near-infrared emission. a,** Schematic illustration of intersystem dual energy transfer between host MADN and radical TTM-TPA in doublet EL devices. **b,** Chemical structures of TTM-TPA, CBP, and MADN. **c,** Absorption coefficient for TTM-TPA in toluene; film PL spectra of CBP neat, MADN neat, and TTM-TPA 3% doped in CBP and MADN. The spectral overlap between TTM-TPA absorption (green squares) with CBP (blue circles) and MADN (red triangles) PL allows singlet-doublet energy transfer.

## High performance near-infrared radical OLEDs

The device structure of the radical OLEDs based on TTM-TPA-doped emitting layer (EML) studied in this work is depicted in Fig. 2a. Using standard OLED design and vacuum deposition, hole injection from ITO/MoO$_3$ was combined with hole transport layers of 1,1-bis[(di-4-tolylamino) phenyl] cyclohexane (TAPC) and 4,4′,4″-tris(carbazol-9-yl)triphenylamine (TCTA). Electron injection was obtained from Al/LiF using a bis-4,6-(3,5-di-3-pyridyl

phenyl)-2-methyl pyrimidine (B3PYMPM) electron transporting layer (ETL). Device characteristics for current density, voltage, EQE, radiance, and EL profile are shown in Fig. 2b-e and summarised in Table 1. The resulting MADN:TTM-TPA OLED gives NIR EL at a peak wavelength of 800 nm with a maximum EQE of 9.6% (Fig. 2b), which is much higher than previous performance limits for reported NIR OLEDs beyond 780 nm peak emission,[1,19] and approximately 60% higher than a maximum EQE of 6.1% obtained in the reference device with CBP:TTM-TPA. Whereas the CBP:TTM-TPA device suffers a large efficiency drop beyond 10 mA cm$^{-2}$, the MADN:TTM-TPA device sustains a relatively high efficiency of 4.2% up to 100 mA cm$^{-2}$. Consequently, the maximum radiance of the MADN:TTM-TPA device reaches 68,000 mW sr$^{-1}$ m$^{-2}$, which is nearly an order of magnitude higher than the 8,100 mW sr$^{-1}$ m$^{-2}$ obtained in the CBP:TTM-TPA device (Fig. 2c).

Fig. 2d shows the NIR doublet EL emission spectra for the devices. Interestingly, in contrast to the PL (Fig. 1c) of CBP:TTM-TPA and MADN:TTM-TPA, no host emission is observed. This indicates that singlet-doublet energy transfer is not the main mechanism at play for EL. Current density-voltage (J-V) plots for devices with and without radical doping are shown in Fig. 2e. Firstly, we observe a steeper J-V gradient for MADN:TTM-TPA versus CBP:TTM-TPA devices. This is consistent with MADN having better electron and hole transporting properties than CBP, as demonstrated using single-carrier device analysis (Supplementary Information S2). Secondly, we find that TTM-TPA doping causes negligible differences between J-V curves for MADN:TTM-TPA and MADN-only devices, whereas a substantially shallower curve is seen in CBP:TTM-TPA versus CBP-only devices. We consider that this indicates radical energy transfer through singlet and triplet channels following exciton formation at host MADN sites in the MADN:TTM-TPA device (Fig. 1a), whereas the J-V characteristics for the CBP-based device suggest the involvement of radical charge trapping.[22] We have tested these devices for stability under constant drive. These devices are exposed to the nitrogen atmosphere between sublimation steps and are operated without encapsulation. Under these conditions, we do find the MADN:TTM-TPA device shows nearly a factor of 10 improved lifetime (to 50% EL) of 58 hr (at 0.1 mA cm$^{-2}$) compared to 7 hr in the CBP:TTM-TPA device (Supplementary Fig. 3), which also presents a substantial increase over previous results.[12,19] The MADN:TTM-TPA device performance sets a new benchmark for stability, which is not generally reported for NIR OLEDs, and higher maximum efficiency and radiance than other reported 780-900 nm devices as summarised in Fig. 2f and Supplementary Table 1.[1,23,24]

**Table 1 | Summary of device performance.**

|  | $V_{on}$[a] (V) | $EQE_{Max}$ (%) | $EQE_{J-1.0}$[b] (%) | $EQE_{J-100.0}$[c] (%) | Radiance$_{Max}$ (mW sr$^{-1}$ m$^{-2}$) | $\lambda_{max}$ (nm) |
|---|---|---|---|---|---|---|
| CBP:TTM-TPA 3% | 2.8 | 6.1 | 4.9 | 1.3 | 8,100 | 820 |
| MADN:TTM-TPA 3% | 2.4 | 9.6 | 7.2 | 4.2 | 68,000 | 800 |

[a] Voltage at 10$^{-1}$ mW sr$^{-1}$ m$^{-2}$, [b] EQE at 1 mA cm$^{-2}$, [c] EQE at 100 mA cm$^{-2}$.

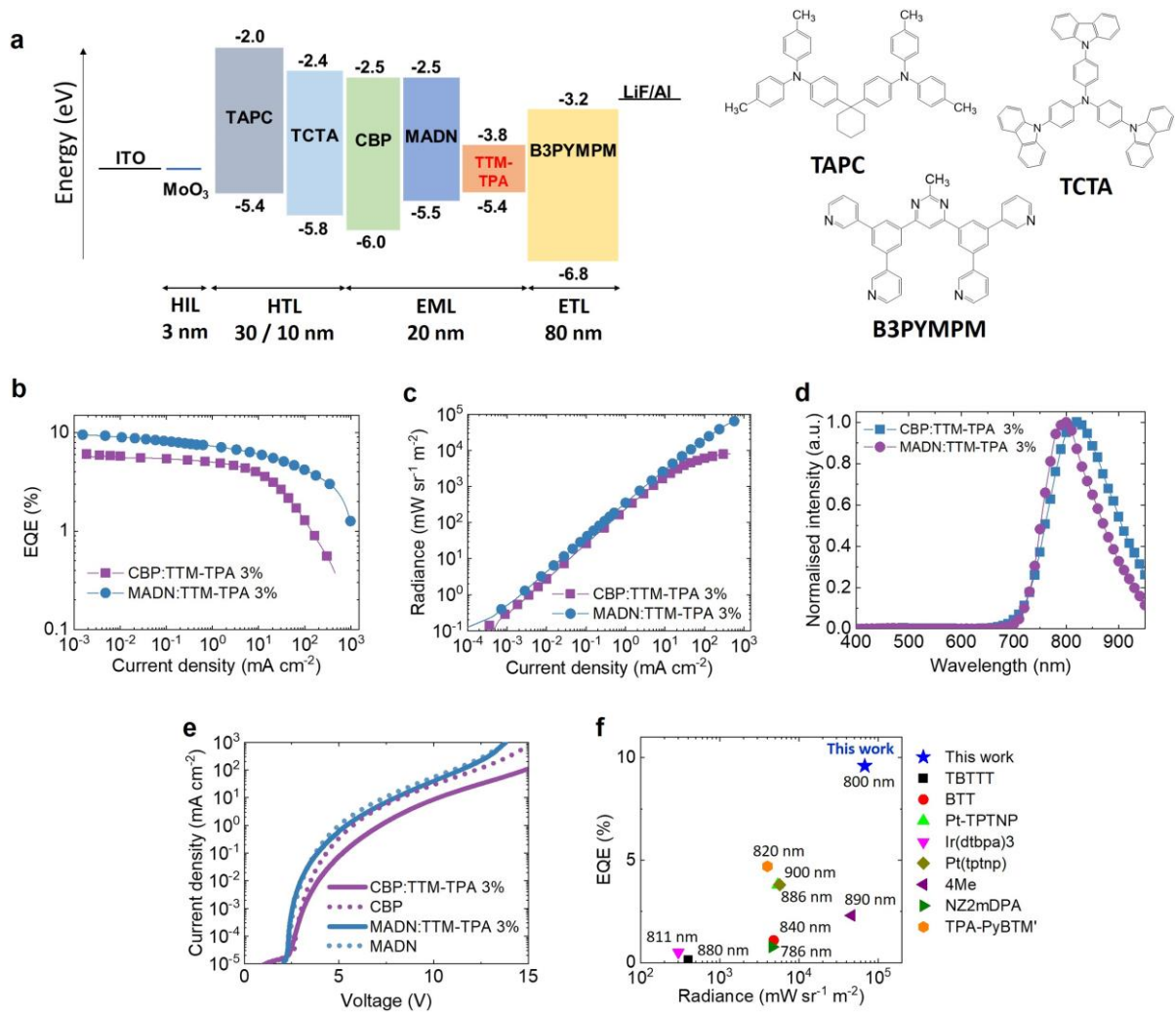

**Fig. 2 | Radical OLED device structure and optoelectronic characterisation. a,** Device structure with energy levels. **b,c,** EQE-current density plots and radiance-current density plots for the devices. **d,** EL spectra at 1 mA cm$^{-2}$. **e,** J-V characteristics for the devices with and without TTM-TPA doping. **f,** Comparison of NIR OLEDs with peak wavelengths between 780 nm and 900 nm regarding maximum EQE and radiance.

The active role of energy transfer in the mechanism for high-performance radical OLEDs was explored by time-resolved optical studies of working devices. On removal of electrical excitation, the transient EL of the MADN-only device shows delayed emission ($\tau$ = 4.3 ± 0.1 µs) that is characteristic of singlet fluorescence following triplet-triplet annihilation (TTA) (blue dotted line, Fig. 3a).[25,26] The CBP-only device shows no delayed emission in transient EL but only prompt decay ($\tau$ = 44 ± 1 ns), which suggests that triplets formed in CBP do not contribute to the overall emission process (purple dotted line, Fig. 3a). As expected from the full device and single-carrier device characteristics (Fig. 2b-e, Supporting Information S2), the CBP:TTM-TPA OLED shows fast prompt doublet fluorescence ($\tau = 150 \pm 3$ ns) where the initial feature of additional emission at ~50-100 ns is attributed to recombination with trapped charges remaining at radical sites from the previous excitation pulse (purple solid line, Fig. 3a).[27–30] In contrast, the MADN:TTM-TPA device exhibits a delayed EL decay of $\tau = 370 \pm 5$ ns (blue solid line, Fig. 3a), which is different from the transient EL profile of the MADN-only device. This supports that the EL mechanism in the MADN:TTM-TPA device involves energy transfer from triplet excitations to emissive radical doublet states without an intermediate TTA process.

Magneto-electroluminescence (MEL) studies were conducted on the MADN:TTM-TPA device and reference devices: MADN-only, CBP-only and CBP:TTM-TPA (Fig. 3b). These studies provide further insights into the EL mechanism by elucidating the effects of magnetic fields on luminescence yield of exciton states in devices.[25,31–36] Here, MEL is defined as: $\text{MEL}(\%) = \frac{\text{EL}(B) - \text{EL}(0)}{\text{EL}(B)}$, where $\text{EL}(B)$ and $\text{EL}(0)$ are the EL intensity in the presence and absence of a magnetic field, $B$, respectively. The CBP-based devices show almost negligible magnetic field dependence of the EL regardless of TTM-TPA doping in CBP-only (purple diamonds) and CBP:TTM-TPA (purple squares) devices in Fig. 3b.

The MADN-based devices with and without TTM-TPA doping are distinguished from the CBP-based devices by positive MEL profiles. The net positive MEL in the MADN-only OLED is attributed to magnetosensitivity of the polaron-pair hyperfine mechanism (positive MEL) that dominates over the dependence from TTA (negative MEL).[25,31,36] The non-identical MEL profiles for the MADN:TTM-TPA device compared to the MADN-only device also imply an EL mechanism without indirect radical energy harvesting by TTA. The broader magnetic field dependence in the MEL profile for the MADN:TTM-TPA OLED is assigned to triplet-doublet energy transfer, where magnetosensitivity originates from larger triplet zero-field splitting interactions (> 10 mT) compared to smaller hyperfine interactions (~1-10 mT) in the polaron-pair mechanism.[31,34]

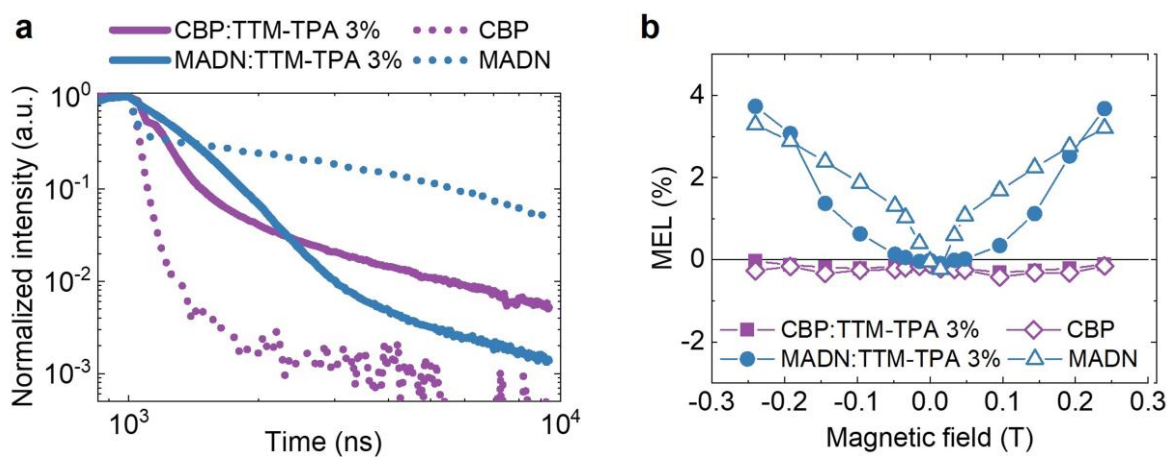

**Fig. 3 | Transient EL and MEL studies of radical OLEDs. a,** Transient EL profiles for CBP and MADN devices with and without TTM-TPA doping. The voltage pulse corresponds to 1 mA cm$^{-2}$, and the off-voltage was -5V for de-trapping the charge carriers after turn-off. In the CBP-based devices (purple), no delayed emission is observed. MADN-based devices (blue) show strong delayed emission features. **b,** MEL for devices studied at 1 mA cm$^{-2}$. The CBP-based devices show negligible MEL while MADN-based devices show positive signatures that reflect energy transfer contributions to the EL mechanism.

**Exciton dynamics and energy transfer**

We performed transient optical spectroscopy studies to investigate the exciton dynamics and the available radical energy transfer pathways depending on $T_1$-$D_1$ energy alignment in these host:radical systems. Picosecond transient absorption (TA) studies were carried out at low fluences to exclude exciton-exciton annihilation. TA under 400 nm pump for host-selective excitation (Fig. 4a) reveals faster decay of MADN excited-state features assigned to $S_1$ excitons in 3% TTM-TPA in MADN films compared to pristine MADN (Supplementary Fig. 7). The $S_1$ decay for MADN:TTM-TPA mirrors a rise in $D_1$ photoinduced absorption from the TTM-TPA component where the timescale for singlet-doublet transfer is rapid ($\tau_{SD}$ = 8 ps). TA studies under radical-selective photoexcitation were performed on CBP:TTM-TPA and MADN:TTM-TPA films with 532 nm excitation (Fig. 4b). Faster decay of $D_1$ excitons is observed in MADN:TTM-TPA, where $|\Delta E_{TD}|$ < 0.1 eV, versus CBP:TTM-TPA, where $|\Delta E_{TD}|$ > 0.8 eV (CBP $T_1$: 2.6 eV).[37] This suggests that radical $D_1$ excitons formed via $S_1 \to D_1$ FRET can transfer energy to closely lying excited triplet $T_1$ states on MADN.

Transient PL studies were conducted on these host:radical films (Fig. 4c,d). Delayed TTM-TPA radical emission is observed in MADN:TTM-TPA film following both host-selective (400 nm, Fig. 4c) and radical-selective (532 nm, Fig. 4d) photoexcitation. In contrast, in CBP:TTM-TPA film, the delayed component contributes less than 1% of all emitted photons under either excitation condition (Supplementary Fig. 9).

The TA and PL dynamics in MADN:TTM-TPA film with low $\Delta E_{TD}$ allow us to conclude that triplet-doublet and doublet-triplet energy transfer pathways are present in this system (Supplementary Information S3). This results in excited-state (re)cycling of triplet and doublet states in a delayed emission mechanism. The temperature dependence of transient PL in MADN:TTM-TPA film shows thermal activation of delayed radical emission under selective radical excitation (Fig. 4e), where the only available processes are doublet luminescence, doublet-triplet and triplet-doublet energy transfer and triplet diffusion. Doublet luminescence is temperature-independent in donor-acceptor TTM radicals.[38] Arrhenius analysis reveals an activation energy of 26.0 ± 1.4 meV (Fig. 4f). This small energy gap is comparable to thermal energy $k_B T$ at room temperature and, therefore, can be efficiently overcome in OLEDs. We assign its origin to diffusion-limited reformation of triplet-radical encounter pairs, as described below.

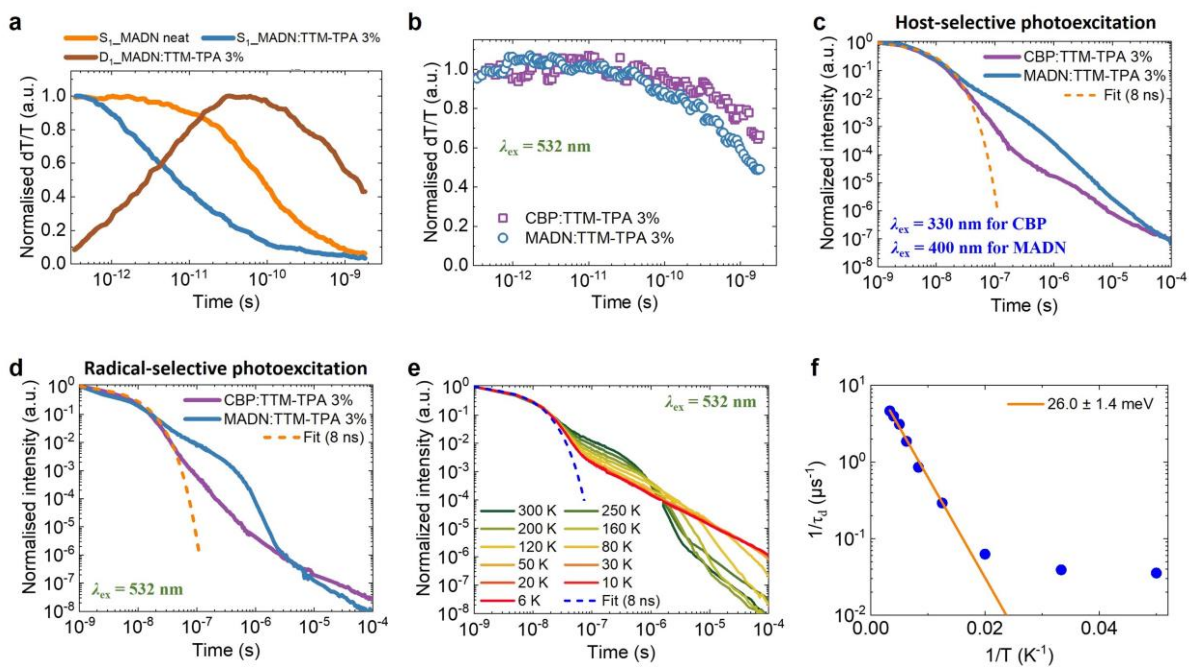

**Fig. 4 | Time-resolved spectroscopy. a,** Excited state singlet ($S_1$) and doublet ($D_1$) population kinetics extracted from transient absorption of neat MADN and MADN:TTM-TPA 3% films under 400 nm excitation. The decay of $S_1$ in the blend and the matching rise of $D_1$ are due to rapid singlet-doublet FRET. **b,** Comparison of $D_1$ population kinetics for CBP:TTM-TPA 3% and MADN:TTM-TPA 3% films under radical-only 532 nm excitation. Faster decay observed in the MADN blend is indicative of doublet-triplet energy transfer. **c,d,** Transient PL profiles averaged over 720-880 nm for radical emission following host-selective (400 nm, 330 nm) and radical-selective (532 nm) excitation. Delayed radical emission is observed in the MADN blend under both host and radical excitation. **e**, Temperature-dependent transient PL profiles of MADN:TTM-TPA 3% excited at 532 nm. Delayed radical emission is faster at elevated temperatures. **f**, Arrhenius plot for the MADN:TTM-TPA 3% system revealing a small activation energy for delayed radical emission.

## Modelling of energy transfer

An amorphous sample made of MADN as the host doped with radical TTM-TPA molecules at a 3.1% m/m concentration was prepared using classical force-field Molecular Dynamics (MD) simulations. After equilibration, a few interacting MADN:TTM-TPA pairs were extracted from the sample, and their ground-state equilibrium geometries were subsequently relaxed at the density functional theory (DFT) level (ωB97X-D/6-31G(d,p)). Vertical excitation energies were computed by resorting to an optimally-tuned screened range-separated hybrid (OT-SRSH) approach (LC-ωhPBE/6-311G(d,p)) within the time-dependent (TD) DFT in the Tamm-Dancoff approximation (TDA) (Supplementary Information S4).[39,40]

In the most stable pair (**dim1**), these calculations yield the first singlet, $S_1$, and triplet, $T_1$, excited states localised on the anthracene core of MADN at 3.17 eV and 2.00 eV, respectively, while the two lowest doublet excited states on TTM-TPA are 1.84 eV ($D_1$) and 2.81 eV ($D_2$) above the ground state. The analysis of the natural transition orbitals (NTOs) for monomers in Supplementary Fig. 11 shows that $D_1$ is an intramolecular charge-transfer (*intra*-$^2$CT) excitation, while $D_2$ has a dominant locally excited ($^2$LE) character on the TTM moiety. The computed excited state energies for all pairs are shown in Fig. 5a. Excitations localised on each fragment show relatively narrow energy distributions, with a slightly larger standard deviation for $D_1$, as expected from its *intra*-$^2$CT character. The calculations also suggest the presence of a much broader distribution of *inter*-$^2$CT excitations, mostly involving transitions from the anthracene core of MADN to the TTM moiety (Supplementary Fig. 13), that are energy-resonant with $D_1$ and $T_1$ and could thus potentially act as mediating states in triplet-doublet energy transfer. The large energy range spanned by these *inter*-$^2$CT states originates from the heterogeneous conformational and electrostatic landscape in amorphous solids.[41,42] These results, *i.e.,* nearly degeneracy of *inter*-$^2$CT with $D_1$ and $T_1$, should be taken with caution since the optimisation of isolated molecular pairs might facilitate the formation of strongly interacting dimers difficult to encounter in the real system. Hence, optimised dimeric models are expected to exhibit shorter intermolecular distances than those in the amorphous material, triggering an overstabilisation of charge-separated states. Indeed, when *inter*-$^2$CT states are directly computed on MD molecular pairs, transition energies are considerably higher (Supplementary Table 7).

Excitation energy transfer (EET) rates for the $S_1$-$D_1$ and $S_1$-$D_2$ processes were computed with the Marcus-Levich-Jortner equation, where all the key parameters (*i.e.,* reorganisation energies, Huang-Rhys factors, electronic couplings and energy differences) were obtained by quantum-chemical calculations. For **dim1**, we calculate an EET time constant from $S_1$ to $D_1$ of 3.0 ps (excluding outliers, an average value of 9 ps is obtained for the investigated pairs, Supplementary Table 9). Despite the smaller energy offset between the states, the corresponding EET time constant from $S_1$ to $D_2$ is significantly longer at 20 ps (average 23 ps) because of reduced Coulomb coupling and smaller oscillator strength associated with $D_2$ compared to $D_1$. We conclude that singlet-doublet energy transfer occurs primarily through the $S_1$-$D_1$ pathway on timescales of a few ps, in excellent agreement with experiment.

We now turn to triplet-doublet energy transfer. A triplet state interacting with a radical can form either an overall doublet or quartet encounter pair in a 1:2 statistical ratio (Fig. 5b).[20] For overall doublet pairs, triplet-doublet energy transfer can occur with spin conservation. The quantum-mechanical coupling between states can take the form of a two-electron exchange integral, as in DET. However, the presence of nearby *inter*-$^2$CT excitations additionally supports a superexchange-mediated mechanism, where the effective coupling is proportional to the product of two, typically much larger, one-electron matrix elements.[38] Building on the pure spin-states of individual fragments, both the direct two-electron and the indirect one-electron electronic couplings were computed for the same pairs as above (Supplementary Information S4). Our calculations show that a direct exchange mechanism provides very slow $T_1$-$D_1$ energy transfer times, with values approaching tens of ns in some pairs. However, superexchange couplings are extremely sensitive to wave function overlap and, therefore, to the dimer geometry and, for some pairs, bring the energy transfer timescales down to tens of ps (which is in the same range as CT-mediated triplet-doublet energy transfer in related covalently-linked radical-chromophore molecules).[38] It is likely that the conversion from the host triplet to the emissive doublet states is limited by diffusion of the triplet excitations within the MADN host. As a first step towards the modelling of triplet diffusion, we computed $T_1$ hopping rates to all nearest neighbours of three randomly selected MADN molecules (Supplementary Information S4). While the values vary over multiple orders of magnitude, the fastest event for the three cases approaches a few tens of ns (Supplementary Table 12), which is typically orders of magnitude slower than the $T_1$-$D_1$ energy transfer. We thus conclude that the thermally activated delayed radical emission is controlled by triplet diffusion within the host, which limits the rate of (re)formation of overall doublet encounter pairs.

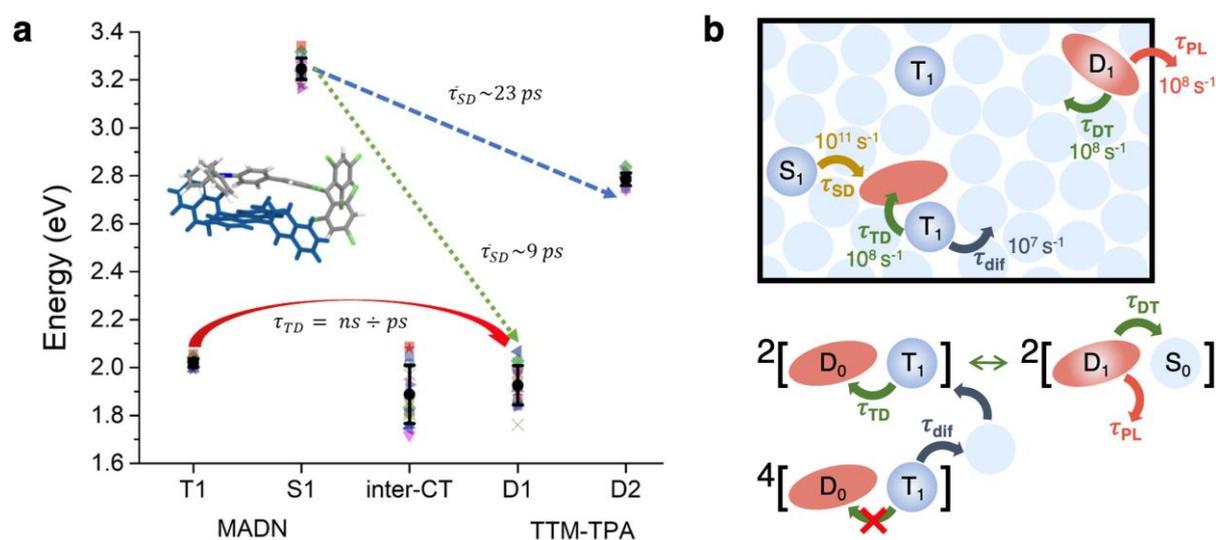

**Fig. 5 | Excited state pathways. a,** Calculated energetic landscape in dimers of MADN and TTM-TPA. The full black circle represents the average adiabatic excitation energies and vertical bars quantify the standard deviation: $T_1$ = 2.02 ± 0.02 eV; $S_1$ = 3.25 ± 0.04 eV; $D_1$ = 1.93 ± 0.08 eV; $D_2$ = 2.79 ± 0.03 eV; $inter$-$^2CT$ = 1.89 ± 0.12 eV. Inset shows the molecular conformation of **dim1**. The computed average lifetime of $\tau_{SD}$ is 23 ps and 9 ps to the $D_2$ and $D_1$ state, respectively. Energy transfer from $T_1$ to $D_1$ occurs in a superexchange-like mechanism mediated by the presence of low-lying $inter$-$^2CT$ states, with a computed overall lifetime $\tau_{TD}$ spanning from tens of ns to tens of ps. **b,** Scheme of exciton pathways and their approximate rates. Triplet excitons form either an overall doublet ($^2[D_0$-$T_1]$) or quartet ($^4[D_0$-$T_1]$) encounter pair when adjacent to a radical site. Reversible energy transfer occurs in the doublet configuration, while quartet pairs separate during triplet diffusion.

## Conclusion

Electrical excitation with a fast charge-transporting host leads to the generation of singlet and triplet exciton states that can be harvested by doublet radicals towards highly efficient NIR EL in OLEDs. Here, the handling of excitations mitigates the energy gap law for non-radiative decay by a design that combines high-energy singlet ($S_1$) and low-energy triplet ($T_1$) excitons of the host with matching to low-energy doublet ($D_1$) excitons of the radical emitter. The principle is demonstrated using the MADN:TTM-TPA combination, which shows rapid singlet-doublet transfer ($\tau = 8$ ps) upon photoexcitation and reversible doublet-triplet cycling with efficient delayed emission ($\tau > 0.16$ µs). The luminescent NIR radical system is implemented in high-performing OLEDs with a maximum EQE of 9.6% for EL at 800 nm that operate to the high maximum radiance of ~68,000 mW sr$^{-1}$ m$^{-2}$, with low efficiency roll-off and enhanced stability. Our design boosts performance in radical-based OLEDs and has broad implications for reducing non-radiative losses in devices beyond light-emitting applications with NIR light.

## Acknowledgements

This work was supported by the Engineering and Physical Sciences Research Council (EPSRC, grant no. EP/M005143/1). E. W. E acknowledges funding from the Royal Society for a University Research Fellowship (URF/R1/201300); and EPSRC grant no. EP/W018519/1. This project has received funding from the ERC under the European Union's Horizon 2020 research and innovation programme (grant agreement number 101020167). H.-H. C. acknowledges George and Lilian Schiff Foundation for Ph.D. studentship funding. P. G. acknowledges the support provided by the Cambridge Trust, George and Lilian Schiff Foundation, Prof. Akshay Rao, and St John's College, Cambridge during the course of the research. The work in Namur and Mons has been funded by the Belgian National Fund for Scientific Research (F.R.S.-FNRS) within the Consortium des Équipements de Calcul Intensif (CÉCI), under Grant No. 2.5020.11, and by the Walloon Region (ZENOBE Tier-1 supercomputer) under Grant No. 1117545. G. L. and Y. O. acknowledge funding from the F.R.S.-FNRS under the grant F.4534.21 (MIS-IMAGINE). D. B. is a FNRS research director. The work at the DIPC was funded by the Spanish Government MICINN (project PID2019-109555GB-I00), the Gipuzkoa Provincial Council (project QUAN-000021-01), the European Union (project NextGenerationEU/PRTR-C17.I1), as well as by the IKUR Strategy under the collaboration agreement between Ikerbasque Foundation and DIPC on behalf of the Department of Education of the Basque Government. D. C and C. T. are thankful for the technical and human support provided by the Donostia International Physics Center (DIPC) Computer Center. C. T. is supported by DIPC and Gipuzkoa's council joint program Women and Science. F. L. are grateful for receiving financial support from the National Natural Science Foundation of China (grant number 51925303).

## Author contributions

H.-H. C. designed a host:guest energy transfer system, fabricated OLEDs, and characterised device performance. H.-H. C., C. C., and P. G. performed steady-state photophysical measurements. H.-H. C. conducted transient EL and MEL measurements. H.-H. C. and S. G. performed transient PL measurements. S. G. performed transient absorption measurements. F. L. contributed to the synthesis of the radical emitter. G. L., S. G., C. T., D. C., Y. O., and D. B. performed quantum-chemical modelling and calculations. H.-H. C., S. G., G. L., D. B., E. W. E., N. C. G., and R. H. F. contributed to writing the manuscript. N. C. G., R. H. F., and E. W. E. supervised the project.

## Additional information

Supplementary information is available in the online version of the paper. Reprints and permissions information is available online at www.nauture.com/reprints. Correspondence and requests for materials should be addressed to N. C. G. (ncg11@cam.ac.uk), R. H. F. (rhf10@cam.ac.uk), and E. W. E. (emrys.evans@swansea.ac.uk). The data that support the plots within this paper are available in the University of Cambridge Repository (URL to be added). Related research results are available from the corresponding authors upon reasonable request.

## Competing interests

The authors declare no competing interests.